\documentclass[aps,prfluids,superscriptaddress]{revtex4-2}

\usepackage{bm}
\usepackage{amsmath,amssymb}
\usepackage{amsfonts}
\usepackage{tensor}
\usepackage{graphicx}

\renewcommand{\d}{\text{d}}

\begin{document}

\title{Stokes equation in a semi-infinite region: generalization of Lamb solution \\ and applications to Marangoni flows}
\author{G. Koleski}
\affiliation{Univ. Bordeaux, CNRS, Laboratoire Ondes et Mati\`ere d'Aquitaine (UMR 5798), F-33400 Talence, France}
\author{T. Bickel}
\email{thomas.bickel@u-bordeaux.fr}

\begin{abstract}
We consider the creeping flow of a Newtonian fluid in a hemispherical  region. In a domain with spherical, or nearly spherical, geometry, the solution of Stokes equation can be expressed as a series of spherical harmonics. However, the original Lamb solution is not complete when the flow is restricted to a semi-infinite space. The general solution in hemispherical geometry is then constructed explicitly.  As an application, we discuss the solutions of Marangoni flows due to a local source at the liquid-air interface.
\end{abstract}

\maketitle

\section{Introduction}

At small length scales,  the creeping flow of a Newtonian fluid is described by the linear Stokes equation. Since the latter is directly related to Laplace equation, it admits a wide variety of analytical solutions that are well documented  and constitute the subject of  classical textbooks~\cite{happelbook,kimbook}. Applications are found in everyday-life, ranging from the actuation of micro-organisms~\cite{laugaRPP2009} to the draining  of liquid foams~\cite{cantatPoF2013} or the wetting dynamics of thin films~\cite{bertinPRL2020}, to name only a few.

Despite these facts, there are still numerous flow problems involving the Stokes equation that remain unresolved. In particular, flows driven by interfacial stresses, such as Marangoni flows, can be especially tenacious~\cite{scriven1960,manikantanJFM2020}.  The first reason for that comes from the presence of the interface itself, which breaks the symmetry in the direction perpendicular to it. The second reason lies in the fact that Marangoni flows are usually coupled to transport phenomena, leading to intricate relations even in the linear regime~\cite{bickelPRFluids2019,bickelEPJE2019}. It was also suggested in recent experimental studies that the presence of surface-active impurities at the interface might induce hydrodynamic instabilities at vanishing Reynolds number~\cite{mizevPoF2005,koleskiPoF2020}. 
There is thus a strong need for a more thorough description of Marangoni flows.

Here, we focus more specifically on the Stokes equation in spherical geometry. The general solution, which was first derived by Lamb~\cite{lambbook}, can be written as a series of Legendre functions. The situation gets more involved when the flow is restricted to a semi-infinite region. In this case, the solution of Lamb is not complete anymore and several authors have pointed out that additional terms have to be considered~\cite{bratukhin,wurgerJFM2014}. The aim of this work is thus to derive in a systematic manner the solution of the Stokes equation in hemispherical geometry. The paper is organized as follows. In Section~\ref{seclamb}, we first remind the main properties of Lamb solution. We then discuss in Section~\ref{secthermocap} the thermocapillary flow due to a point-like heat source at the interface.  This simple example shows unambiguously that Lamb solution happens to be incomplete when considering a hemispherical domain. The general solution is then constructed in Section~\ref{secgeneralsol}, and  some applications to thermocapillary flows are  given in Section~\ref{secappli}. The main results are finally summarized in Section~\ref{conclusion}.

\section{Stokes equation in spherical coordinates}
\label{seclamb}

\subsection{Lamb solution}

We consider the viscous flow of an  incompressible fluid in a domain with spherical, or nearly spherical, geometry. We note $r$ the radial coordinate, $\theta$ the polar angle ($0 \leq \theta \leq \pi$) and $\varphi$  the azimuthal angle ($0 \leq \varphi < 2 \pi$). In the limit of vanishing Reynolds number, the velocity and pressure fields are governed by the Stokes equation
\begin{equation}
\eta \nabla^2 \mathbf{v} = \bm{\nabla} p \ , 
\label{stokes}
\end{equation}
together with the continuity equation
\begin{equation}
 \bm{\nabla} \cdot \mathbf{v} = 0 \ .
\label{incomp}
\end{equation}
Due to the linearity of Equations~(\ref{stokes}) and~(\ref{incomp}), the general solution can be written as the sum of a homogeneous and a particular solution, $\mathbf{v} =\mathbf{v}_H+ \mathbf{v}_P$. 
\begin{itemize}
\item The homogeneous solution $\mathbf{v}_H$  satisfies the equations $\nabla^2 \mathbf{v}_H = \bm{0}$ and $\bm{\nabla}\cdot \mathbf{v}_H = 0$. It can thus be expressed as the linear combination of a potential and a toroidal field 
\begin{equation}
\mathbf{v}_H = \bm{\nabla}\Phi + \bm{\nabla}\times \left( \mathbf{r}  \chi  \right) \ ,
\end{equation}
where $\Phi$ and $\chi$ are both harmonic functions, $\nabla^2 \Phi = \nabla^2 \chi =0$. 
\item The particular solution $\mathbf{v}_P$ is  related to the theory of harmonic functions as well. Taking the divergence of the equation $\nabla^2 \mathbf{v}_p = \bm{\nabla} p$ together with $\bm{\nabla}\cdot \mathbf{v}_p = 0$, it is straightforward to show  that the pressure field  satisfies  Laplace equation, $\nabla^2 p = 0$.
\end{itemize}
In spherical coordinates, the solution of  Laplace equation $\nabla^2 \psi = 0$ can be expressed as a series of spherical harmonics. It can be written quite generally as $\psi (r,\theta,\varphi)  =\sum_{l=-\infty}^{\infty} \psi_l(r,\theta,\varphi)$, with
\begin{equation}
\psi_l(r,\theta,\varphi) =  \sum_{m=-l}^{l}  A_{(lm)} r^{l}  Y_l^m(\theta,\varphi)  \ .
\label{decomposition}
\end{equation}
The spherical harmonics of degree~$l$ and order $m$ are defined (up to a numerical prefactor) by
\begin{equation} 
Y_l^m(\theta,\varphi)  = P_l^m(\cos \theta)  e^{im\varphi}  \ ,
\end{equation}
where the associated Legendre polynomials can be expressed for instance by Rodrigues formula
\begin{equation}
P_l^m(x) = \frac{(-1)^{l+m}}{2^l l!}\left(1-x^2\right)^{\vert m\vert /2} \frac{\d^{l+\vert m\vert}}{\d x^{l+\vert m\vert}} \left( 1 -x^2  \right)^l \quad \text{for} \quad l \geq 0 \ . 
\label{defasslegendre}
\end{equation}
The latter relation states in particular that   $Y_l^m =0$ if $\vert m \vert >l$.  Remark that the definition of $Y_l^m$ can be extended to negative degrees thanks to the equality $Y_l^m=Y_{-l-1}^m$. An essential property of spherical harmonics is that they form an orthogonal basis for scalar functions on the sphere surface. 

Coming back to the Stokes equation,  we can implement then same decomposition as in Equation~(\ref{decomposition}) for the potential, toroidal and pressure fields, respectively.  It is then possible to show that 
\begin{equation}
\mathbf{v}(\mathbf{r}) = \sum_{l=-\infty}^{\infty} \left[ \frac{(l+3)r^2  \bm{\nabla}p_l}{2\eta (l+1)(2l+3)} -\frac{  \mathbf{r} l p_l}{\eta (l+1)(2l+3)}  + \bm{\nabla}\Phi_l + \bm{\nabla}\times \left( \mathbf{r}  \chi_l \right) \right] \ .
\label{lamb}
\end{equation}
This expression is usually named after Lamb~\cite{lambbook}. 
It is therefore a series in powers (for $l>0$)  or inverse powers (for $l<0$) of the distance~$r$. 
As a matter of fact, the convention $l \leftrightarrow -l-1$ is often assumed for the exterior flows in order to avoid negative indices.
The components of the velocity $\mathbf{v}=v_r \boldsymbol{e}_r+ v_{\theta} \boldsymbol{e}_{\theta}+ v_{\varphi} \boldsymbol{e}_{\varphi}$ are finally obtained by applying the differential operators in Equation~(\ref{lamb}). 

\subsection{Explicit expression for exterior flows}

The Lamb solution is relevant to describe the motion of the fluid inside  or outside  a region with spherical symmetry.  In this work,  we focus specifically on exterior flows and assume that all fields vanish when $r \to \infty$. The pressure can then be conveniently expressed as
\begin{equation}
p =  \eta \sum_{l\geq1} \sum_{m}    \frac{ \pi_{(lm)}}{r^{l+1}} \frac{2 (2l-1)}{(l+1)} Y_l^m  \ ,
\label{pext}
\end{equation}
with $-l \leq m \leq l$.
The  components of the velocity $\mathbf{v}= v_r \mathbf{e}_r + v_{\theta} \mathbf{e}_{\theta} + v_{\varphi} \mathbf{e}_{\varphi}$  are given by~\cite{happelbook,kimbook} 
\begin{subequations}
\label{vlamb}
\begin{align}
& v_r =  \sum_{l\geq1} \sum_{m}   \frac{\pi_{(lm)}}{r^{l}}  Y_l^m + \sum_{l\geq3} \sum_{m}   \frac{ \rho_{(l-2,m)}}{r^{l}} Y_{l-2}^m     \ , \label{vrext} \\
& v_{\theta} =  \sum_{l\geq1} \sum_{m}  \frac{\pi_{(lm)}}{r^{l}}  \frac{(l-2)}{l(l+1)}   s  \partial_c Y_l^m +  \sum_{l\geq3} \sum_{m}  \frac{ \rho_{(l-2,m)}}{r^{l}}  \frac{1}{(l-1)} s \partial_c Y_{l-2}^m
+  \sum_{l\geq2} \sum_{m}    \frac{\sigma_{(l-1,m)}}{r^l} \frac{im}{s}   Y_{l-1}^m   \ ,  \label{vthetaext} \\
& v_{\varphi} = -  \sum_{l\geq1} \sum_{m}   \frac{\pi_{(lm)}}{r^{l}}    \frac{(l-2)}{l(l+1)}     \frac{im}{s} Y_l^m -  \sum_{l\geq3} \sum_{m}  \frac{ \rho_{(l-2,m)}}{r^{l}}  \frac{1}{(l-1)} \frac{im}{s} Y_{l-2}^m
+ \sum_{l\geq2} \sum_{m}  \frac{\sigma_{(l-1,m)}}{r^l}   s \partial_c \mathcal{Y}_{l-1}^m     \ , \label{vphiext}
\end{align}
\end{subequations}
where we have introduced $c= \cos \theta$ and $s=\sin \theta$. Note that our notation differs slightly from that usually found in the literature, since we anticipate the generalization to a semi-infinite region. 

As an illustration, let us consider the axisymmetric solutions that decays from the origin as the inverse distance. The radial and polar components of velocity field then read  $v_r(r,\theta) = \Phi^{(0)}_r(\theta)/r$ and $v_{\theta}(r,\theta) = \Phi^{(0)}_{\theta}(\theta)/r$, where the functions $\Phi^{(0)}_r$ and $\Phi^{(0)}_{\theta}$ are given by
\begin{equation}
\Phi^{(0)}_r (\theta)=    \pi_{(10)} P_1^0 \ , \quad \text{and} \quad \Phi^{(0)}_{\theta} (\theta) = -  \frac{ \pi_{(10)}}{2}  s \partial_c  P_1^0 \ ,
\label{lambfirst}
\end{equation}
with $ P_1^0 = \cos \theta$ and  $s \partial_c P_1^0 = \sin \theta$.
The integration constant $\pi_{(10)}$ has then to be set by enforcing the relevant boundary conditions.

\section{A first encounter with hemispheric Legendre functions}
\label{secthermocap}

So far, we have made no restriction regarding the polar angle~$\theta$. The issue gets more involved, however, when it is limited to part of the interval $[0,\pi]$.   Such a situation typically occurs when the fluid is bounded by a flat surface or interface. This encompasses a large number of physical realizations, ranging from a droplet placed on a solid surface to the motion of a camphor scrap at the water-air interface. 
In a semi-infinite domain, the solution of Lamb is not complete anymore and several authors have recently derived additional terms to complement Equation~(\ref{lamb}). To illustrate our point, we first discuss a classic problem of fluid mechanics: the steady-state Marangoni flow due to a local heat source at the liquid-air interface~\cite{bratukhin}. 

\begin{figure}
\includegraphics[width=0.7\textwidth]{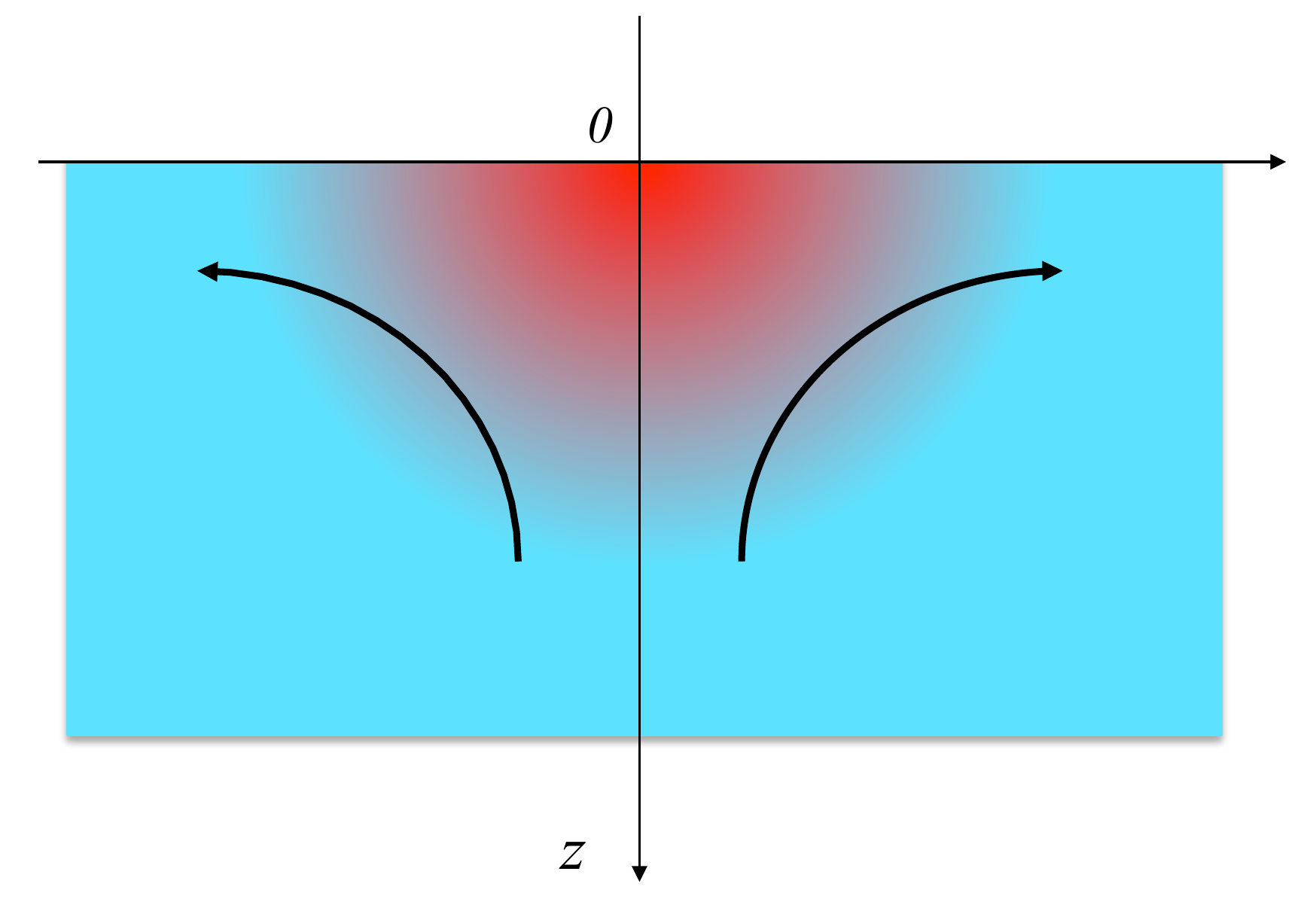}
\caption{Schematic representation of the system. The temperature gradient induces horizontal stresses that drive the fluid from the hot to the cold areas of the interface. The $z$-axis is oriented downward.}
\label{schema}
\end{figure}

\subsection{Thermocapillary flow due to a point source}

We study the stationary heat flow induced by a source located at $z=0$. This can be achieved for instance with a colloidal particle heated by a laser beam~\cite{girotLangmuir2016}. 
Assuming that the size of the source is small compared to the other relevant length-scales, heat may be regarded as emerging from a point source  $q(\mathbf{r})= Q\delta (\mathbf{r})$, with $Q$ the injected power ($[Q]=W$), in a semi-infinite liquid bounded by a flat  interface. The $z$-axis is perpendicular to the interface and oriented downward, the liquid phase corresponding to $z>0$. The system is schematically shown in Figure~\ref{schema}.
We assume that the temperature variations are small enough so that both the viscosity and the fluid density can be regarded as constant.

The discussion is restricted to the linearized version of the transport equations, \textit{i.e.}, we focus on the limits of vanishing Reynolds and P\'eclet numbers.
The temperature field is then solution of the heat equation
\begin{equation}
\nabla^2 T = - \frac{Q}{\kappa} \delta (\mathbf{r}) \ ,
\label{heat}
\end{equation}
with $\kappa$ the thermal conductivity ($[\kappa]=\text{W}\cdot \text{m}^{-1}\cdot\text{K}^{-1}$), and $\delta$ the Dirac delta function ($[\delta(\mathbf{r})] = \text{m}^{-3}$ in 3D). The velocity field satisfies the incompressible Stokes Equations~(\ref{stokes}) and~(\ref{incomp}). 
Regarding the boundary conditions, it is assumed that  there is neither mass nor heat flux across the interface 
\begin{equation}
 \mathbf{v}  \cdot   \boldsymbol{e}_z = 0  \quad \text{and} \quad \bm{\nabla}  T \cdot   \boldsymbol{e}_z = 0 \quad \text{at} \quad z=0 \ ,
 \label{bc0}
\end{equation}
whereas both the velocity  and the temperature return to their equilibrium values far away from the disturbance
\begin{equation}
\lim_{r \to \infty}  \mathbf{v}( \mathbf{r})  = \bm{0}  \quad \text{and} \quad \lim_{r \to \infty}  T(\mathbf{r})  = T_0 \ .
\label{bcinfinity}
\end{equation}
Finally, since the surface tension is usually a decreasing function of the temperature, the actuation of the fluid arises from stresses at the interface. This is expressed by the Marangoni boundary condition, that relates the shear stress and the gradient of interfacial tension
\begin{equation}
\left(   \boldsymbol{1} -   \mathbf{e}_z \mathbf{e}_z \right)  \cdot  \left( \boldsymbol{\sigma} \cdot \mathbf{e}_z - \bm{\nabla} \gamma  \right)\Big\vert_{z=0} = 0   \ , 
\label{bcmar}
\end{equation}
with $\boldsymbol{\sigma}$ the hydrodynamic stress tensor [whose cartesian components are $\sigma_{ij}=-p\delta_{ij} + \eta \left( \partial_i v_j + \partial_j v_i \right)$].
For moderate deviations with respect to the equilibrium temperature, one can assume a linearized relationship for the surface tension: $\gamma(T) = \gamma_0 -\gamma_T (T-T_0)$,
with $\gamma_0=\gamma(T_0)$.  The  coefficient $\gamma_T = \vert \partial \gamma / \partial T \vert$   characterizes the rate of change of  surface tension with respect to temperature. 

\subsection{Solution of the transport equations}

The mathematical problem defined by Equations~(\ref{heat})--(\ref{bcmar}) is  solved in the semi-infinite region $z>0$.
First, the temperature field can be derived by analogy with electrostatics. Indeed, Equation~(\ref{heat}) is equivalent to the Poisson equation for the electrostatic potential due to a point charge. One therefore gets
\begin{equation}
T (\mathbf{r}) = T_0 + \frac{Q}{2\pi \kappa r} \ .
\label{tempfield}
\end{equation}
Note that the numerical coefficient differs from the usual value ($2 \pi$ \textit{vs.} $4 \pi$) since heat diffusion is restricted to a half-space. Note also that this solution is only valid down to a cut-off distance set by the size of the physical heat source.

Next, Equation~(\ref{bcmar})  suggests that the velocity fields follows the same power-law as the temperature. We can therefore search for solutions of the Stokes Equations~(\ref{stokes})--(\ref{incomp}) in the form $v_r(r,\theta) = \Phi_r(\theta)/r$ and $v_{\theta}(r,\theta) = \Phi_{\theta}(\theta)/r$. After some algebra, one obtains
\begin{equation}
\Phi_r(\theta) = \frac{Q\gamma_T}{4\pi \kappa \eta} (1- 2\cos \theta)   \ , \quad \text{and} \quad \Phi_{\theta}(\theta) = \frac{Q\gamma_T}{4\pi \kappa \eta}  \frac{\cos \theta \sin \theta}{1+\cos \theta} \ .
\label{bratukhin}
\end{equation}
At first sight, the connexion between the $\theta$-dependance in Equation~(\ref{bratukhin}) and the generic Lamb solution Equation~(\ref{lambfirst}) is not striking. Still, it can be noticed that the function $\Phi_r(\theta)$ can also be written as 
\begin{equation}
\Phi_r(\theta) = \alpha P_1^0(c) -\frac{\alpha}{2} \ ,
\label{fc}
\end{equation}
 with $P_1^0(c)=c=\cos \theta$ and $\alpha=-Q\gamma_T/(2\pi \kappa \eta)$. Similarly, $\Phi_{\theta}(\theta)$ can be re-expressed as
\begin{equation}
\Phi_{\theta}(\theta) = -\frac{\alpha}{2}  s \partial_c  P_1^0 + \frac{\alpha}{2} \sqrt{\frac{1-c}{1+c}}  \ , 
\label{gc}
\end{equation}
 with $ s \partial_c  P_1^0=s=\sin \theta$. In both expressions, the first term involves an associated Legendre function (or its derivative) with $l=1$ and $m=0$, as expected from  Lamb solution~(\ref{lambfirst}). The second term, which is not present in Equation~(\ref{lambfirst}), is more puzzling.
Let us focus for instance on the expression~(\ref{gc}) for $\Phi_{\theta}$: although the second term is well defined in the upper half-space $0 \leq c \leq 1$, it is obviously singular in the limit $c \to -1$. Since the original Lamb solution only involves functions that are regular in the full range $-1 \leq c \leq 1$, this contribution is necessarily missing from the general solution.
As we shall see in the following, the  additional terms in Equations~(\ref{fc}) and~(\ref{gc}) can actually be defined as \textit{hemispheric Legendre functions}.

\section{Laplace and Stokes equations in a semi-infinite region}
\label{secgeneralsol}

As argued in Section~\ref{seclamb}, the solution of  the Stokes equation is closely related to the theory of harmonic functions. 
In this section, we first focus on  Laplace equation in order to introduce generalized spherical harmonics. The derivation follows closely that of Reference~\cite{seabornbook}. We then discuss how to extend the solution of Lamb in a semi-infinite domain.

\subsection{Method of separation of variables applied to Laplace equation}

Let us consider the Laplace equation $\nabla^2 \psi(\mathbf{r})=0$ satisfied by a real scalar function $\psi$. The position vector~$\mathbf{r} = (r, \theta,\varphi)$ is restricted to the upper  half-space $z \geq 0$, which corresponds  to the interval $0 \leq \theta \leq \pi/2$.  Results pertaining to the lower half-space $z \leq 0$ can be deduced in a straightforward manner thanks to the symmetry $\theta \leftrightarrow \pi - \theta$.

In spherical polar coordinates, the Laplace equation reads
\begin{equation}
\nabla^2 \psi = \frac{1}{r^2} \frac{\partial}{\partial r} \left( r^2 \frac{\partial  \psi}{\partial r} \right)  + \frac{1}{r^2\sin \theta} \frac{\partial }{\partial \theta} \left( \sin \theta \frac{\partial  \psi}{\partial \theta} \right) + \frac{1}{r^2 \sin^2\theta } \frac{\partial^2  \psi}{\partial \varphi^2} = 0 \ .
\label{laplace}
\end{equation}
This equation can be solved using the method of separation of variables. To this aim, we seek a general solution of the form $\psi(r,\theta,\varphi)= f(r) g(\theta) h(\varphi)$, where $f$, $g$ and $h$ are functions of a single variable. Equation~(\ref{laplace}) then leads to a set of ordinary differential equations
\begin{align}
& \frac{\d}{\d r} \left( r^2 \frac{\d f}{\d r} \right) - \lambda f = 0  \ ,  \label{laplacef} \\
& \frac{1}{\sin \theta} \frac{\d }{\d \theta} \left( \sin \theta \frac{\d  g}{\d \theta} \right) + \left( \lambda - \frac{\mu}{\sin^2 \theta} \right) g = 0  \ , \label{laplaceg} \\
& \frac{\d^2 h}{\d \varphi^2} + \mu h = 0 \ , \label{laplaceh}
\end{align}
where $\lambda$ and $\mu$ are two (yet unknown) constants. 

We first consider the radial Equation~(\ref{laplacef}). Without loss of generality, we can set $\lambda = l(l+1)$. Searching for power-law solutions, one readily gets
\begin{equation}
f(r) = A r^{l} + B r^{-l-1} \ ,
\label{solf}
\end{equation}
 where $A$ and $B$ are two integration constants. In the particular case $l=-1/2$, the general solution~(\ref{solf}) has to be replaced by $f(r)=Ar^{-1/2}+Br^{-1/2}\ln r$.

We next focus on the azimuthal Equation~(\ref{laplaceh}). On physical grounds, one requires the function $h$ to be single-valued. Since $(r,\theta,\varphi)$ and $(r,\theta,\varphi+2 \pi)$ represents the same location in space, the condition $h(\varphi + 2 \pi) = h(\varphi)$ demands that $\mu = m^2$, with $m$ an integer. We write the solution of Equation~(\ref{laplaceh})  as
\begin{equation}
h(\varphi) = e^{im\varphi} \ , \quad \text{with} \quad m\in \mathbb{Z} \ .
\end{equation}

\subsection{Hemispheric Legendre functions}

Regarding Equation~(\ref{laplaceg}), it can be recast  in a more usual form by setting $c = \cos \theta$. Since  we focus on the semi-infinite region $0 \leq \theta \leq \pi/2$, the variable $c$ is always positive: $0 \leq c \leq 1$. Equation~(\ref{laplaceg}) is then equivalent to Legendre differential equation
\begin{equation}
(1-c^2) \frac{\d^2 g}{\d c^2}-2c\frac{\d g}{\d c}+\left[ l(l+1) -\frac{m^2}{1-c^2} \right] g(c) = 0 \ .
\label{asslegendreeq}
\end{equation}

In the situations commonly encountered in physics, \textit{e.g.} in quantum mechanics or in fluid mechanics, one is interested in solutions of Equation~(\ref{asslegendreeq}) that are regular over the entire interval $c \in [-1,1]$. These standard solutions  are  the well-known associated Legendre functions $P_l^m(c)$ defined in Section~\ref{seclamb}. 
In this work, however, the regularity condition  is only required in the semi-infinite region $\mathcal{I}^+ = [0,1]$. Additional solutions, which can be singular on the  complementary interval~$\mathcal{I}^-=[-1,0]$, are therefore permitted.  The essential point is that the extra singularities that may exist in~$\mathcal{I}^-$ are physically irrelevant since we exclusively focus on the interval~$\mathcal{I}^+$.
Although this issue is fairly well established in classical physics, \textit{e.g.} in the context of the sharp point effect in electrostatics~\cite{jacksonbook}, we find it worth to remind the main steps of the derivation --- see Appendix~\ref{appgauss}.
The general solution of Equation~(\ref{asslegendreeq}) is therefore $g(c) = \mathcal{P}_l^m(c)$, where the hemispheric Legendre functions are defined as
\begin{equation} 
\mathcal{P}_l^m(c)= \left( \frac{1-c}{1+c} \right)^{\vert m \vert /2} F\left(-l,l+1;1+\vert m \vert;\frac{1-c}{2}\right) \ , \quad c\in[0,1]  \ ,
\label{defplm}
\end{equation}
where the hypergeometric function $F$ is given by the series 
\begin{equation} 
F(\alpha,\beta;\gamma;x) = \sum_{n=0}^{\infty} \frac{(\alpha)_n(\beta)_n}{(\gamma)_n} \frac{x^n}{n!} \ ,
\label{defhypergeo}
\end{equation}
with $(q)_n$  the Pochhammer symbol: $(q)_n=q(q+1)\ldots (q+n-1)$ for $n >0$, and $(q)_0=1$.
The main properties of the hypergeometric function are listed in Appendix~\ref{appgauss}.

In Equation~(\ref{defplm}), the degree $l$ can assume any real value. 
Working in a semi-infinite region thus implies a rather unconventional definition of the associated Legendre functions.  For  $0 \leq \vert m \vert \leq l$, they coincide (up to a numerical prefactor) with the ``usual''  functions $P_l^m(c)$ defined in~(\ref{defasslegendre}) --- see Table~\ref{tablelegendre}.
However, since the series~(\ref{defhypergeo}) actually converges for all $c\in [0,1]$, there is no restriction regarding the order $m$ anymore. Solutions with $l<\vert m \vert$ do exist as well, although this is normally forbidden in $\mathbb{R}^3$. This assertion is of paramount importance since these additional terms play a central role to account for interfacial phenomena such as the Marangoni effect.
It can be noticed that the generalized Legendre functions with $\vert m \vert >l$ are proportional to $(1+c)^{-\vert m \vert /2}$. These functions are indeed regular in the semi-infinite region $0 \leq c \leq 1$, but would be otherwise  singular on the whole interval $-1 \leq c \leq 1$. 
To illustrate our point,  the first hemispheric Legendre functions are listed in Table~\ref{tablelegendre}.

\begin{table}[h]
\caption{Expression of the first hemispheric Legendre functions $\mathcal{P}_l^m(c)$ for $c>0$.}
\label{tablelegendre}
\begin{center}
{\renewcommand{\arraystretch}{2.5}
\begin{tabular}{| c | c | c | c |  c | }
\hline
 & $m=0$ & $m=1$ & $m=2$ & $m=3$  \\
\hline
$l=0$ &  $1$ & $\displaystyle{\left(\frac{1-c}{1+c}\right)^{1/2}}$ & $\displaystyle{\frac{1-c}{1+c}}$ & $\displaystyle{\left(\frac{1-c}{1+c}\right)^{3/2}}$ \\
\hline
$l=1$ & $c$ & $\displaystyle{\frac{1}{2}} \left(1-c^2\right)^{1/2}$ & $\displaystyle{\frac{1}{3}(2+c)\left(\frac{1-c}{1+c}\right)}$ &$\displaystyle{\frac{1}{4}(3+c)\left(\frac{1-c}{1+c}\right)^{3/2}}$ \\
\hline
$l=2$ &$\displaystyle{\frac{1}{2}(3c^2 - 1)}$ &  $\displaystyle{\frac{1}{2}}c \left(1-c^2\right)^{1/2}$   &  $\displaystyle{\frac{1}{4} (1-c^2)}$ &$\displaystyle{\frac{1}{20} (8+9c+3c^2)\left(\frac{1-c}{1+c}\right)^{3/2}}$ \\
\hline
\end{tabular}}
\end{center}
\end{table}

Finally,  the general solution of Laplace equation $\nabla^2 \psi(\mathbf{r})=0$  can be written in spherical coordinates as 
\begin{equation} 
\psi(r,\theta,\varphi) = \int_{-\infty}^{\infty} \d l \sum_{m=-\infty}^{\infty} A_{(lm)} r^{l}  \mathcal{Y}_l^m(\theta,\varphi)  \ .
\label{decompgeneral}
\end{equation}
Here, we define the hemispheric spherical harmonics
\begin{equation} 
\mathcal{Y}_l^m(\theta,\varphi)  = \mathcal{P}_l^m(c) e^{im\varphi} \ ,
\end{equation}
with $\mathcal{P}_l^m(c)$ given in Equation~(\ref{defplm}).
The first difference between the solution~(\ref{decomposition}) in  $\mathbb{R}^3$ and Equation~(\ref{decompgeneral}) in the semi-infinite region $0\leq c \leq 1$ is that the sum now extend to any value of $m$, be it smaller or larger than the degree~$l$. The second difference is that the degree~$l$ varies continuously, the sum being replaced by an integral.
Note also that the functions $\mathcal{Y}_l^m$ (or $\mathcal{P}_l^m$) are in general not orthogonal on the half-sphere $0 \leq c \leq 1$ (although orthogonality is preserved regarding the azimuthal angle $\varphi$) .

\subsection{Generalized solution for exterior flows}

We now have all the elements to extend the Lamb solution to a hemispheric region. 
Two strategies are possible. The first is to directly transpose the solution given by
 Equations~(\ref{pext})--(\ref{vlamb}), provided the correspondence $Y_l^m \ \Leftrightarrow  \  \mathcal{Y}_l^m$ and
\begin{equation} 
\sum_{l\geq l_{0}} \sum_{m=-l}^l   \ldots \quad  \Leftrightarrow \quad  \int_{0}^{\infty} \d l \sum_{m=-\infty}^{\infty} \ldots  \ ,
\label{correspondence}
\end{equation}
with $l_0=1$, 2 or 3. As we shall see, this approach works for all but a few terms of the series. The second strategy consists in solving explicitly the Stokes equation. This gives of course the same results, but with the advantage that singular contributions can be handle consistently. Details regarding the algebra are given in Appendix~\ref{appl0}. 
Focusing as previously on the exterior solution, we find that the pressure and the components of the velocity can be written as
\begin{subequations}
\label{componentgen}
\begin{align}
& p (r, \theta , \varphi)  = \int_{0}^{\infty} \d l \sum_{m=-\infty}^{\infty}  \frac{1}{r^{l+1}} p_{lm}(c) e^{im\varphi}  \ , \\
& v_r (r, \theta , \varphi) =  \int_{0}^{\infty} \d l \sum_{m=-\infty}^{\infty} \frac{1}{r^l} u_{lm}(c) e^{im\varphi} \ , \\
& v_{\theta} (r, \theta , \varphi) = \int_{0}^{\infty} \d l \sum_{m=-\infty}^{\infty} \frac{1}{r^l} v_{lm}(c) e^{im\varphi} \ ,  \\
& v_{\varphi} (r, \theta , \varphi) = \int_{0}^{\infty} \d l \sum_{m=-\infty}^{\infty}  \frac{1}{r^l} w_{lm}(c) e^{im\varphi} \ , 
\end{align}
\end{subequations}
where we get
\begin{subequations}
\label{velocitygen}
\begin{align}
& p_{lm}(c)  = \pi_{(lm)} \frac{2 (2l-1)}{(l+1)} \mathcal{P}_l^m  \ . \label{pgen} \\
& u_{lm}(c) = \pi_{(lm)} \mathcal{P}_l^m +  \rho_{(l-2,m)} \mathcal{P}_{l-2}^m  \ ,  \label{vrgen}\\
& v_{lm}(c) =  \pi_{(lm)} \frac{(l-2)}{l(l+1)}   s  \partial_c  \mathcal{P}_l^m +  
 \rho_{(l-2,m)}  \frac{1}{(l-1)} s \partial_c \mathcal{P}_{l-2}^m +  \sigma_{(l-1,m)}   \frac{im}{s} \mathcal{P}_{l-1}^m  \ ,  \label{vthetagen} \\
& w_{lm}(c) = -  \pi_{(lm)}  \frac{(l-2)}{l(l+1)}   \frac{im}{s}  \mathcal{P}_l^m
-  \rho_{(l-2,m)}   \frac{1}{(l-1)} \frac{im}{s} \mathcal{P}_{l-2}^m+ \sigma_{(l-1,m)}  s \partial_c \mathcal{P}_{l-1}^m \ , \label{vphigen} 
\end{align}
\end{subequations}
Equations~(\ref{componentgen}) and~(\ref{velocitygen}) are the main outcomes of this work. They express the exterior solution of a Stokes flow as a straightforward generalization of the original Lamb solution. Still, there are a few terms in the series that are singular and should therefore be handled separately.

\subsubsection{Polar and azimuthal components of the velocity for $l=1$ and $m \neq 0$.}

It can be noticed in Equations~(\ref{vthetagen}) and~(\ref{vphigen}) that both the polar and azimuthal components of the velocity are not properly defined for $l=1$.
To regularized the expressions, we need to solve explicitly the incompressible Stokes equations. This can be achieved by appealing the recurrence relations of Legendre functions. The main steps of the derivation are given in Appendix~\ref{appl0}, so that we directly give the results
\begin{subequations}
\label{vl1gen}
\begin{align}
& v_{1m}(c) = -  \frac{\pi_{(1m)}}{2}   s  \partial_c  \mathcal{P}_1^m +  
(1+\vert m \vert ) \rho_{(-1,m)} \frac{\mathcal{P}_{1}^m}{s} +   \sigma_{(0,m)}   \frac{im}{s} \mathcal{P}_{0}^m  \ ,  \label{vthetal1gen} \\
& w_{1m}(c) =  \frac{\pi_{(1m)}}{2}    \frac{\partial_{\varphi} \mathcal{P}_l^m}{s} 
+ \frac{is}{m}   \rho_{(-1,m)}  \left[ \mathcal{P}_{0}^m - (1+ \vert m \vert) \partial_c \mathcal{P}_{1}^m \right] + \sigma_{(0,m)}  s \partial_c \mathcal{P}_{0}^m \ . \label{vphil1gen} 
\end{align}
\end{subequations}

\subsubsection{Polar component of the velocity for $l=1$ and $m = 0$.}

As a matter of fact, Equation~(\ref{vthetal1gen}) is still singular when the order $m$ is equal to zero. To remove this very last singularity, we 
proceed as previously and solve directly the differential equations --- see Appendix~\ref{appl0}. We obtain the explicit expression
\begin{equation}
v_{10}(c) = -  \frac{\pi_{(10)}}{2} s -  \rho_{(-1,0)} \sqrt{ \frac{1-c}{1+c} }  \ ,  \label{vthetal10gen}
\end{equation}
Regarding the azimuthal component for $m=0$, one can deduce from Equation~(\ref{vphigen}) that $w_{l0}(\theta , \varphi)=\sigma_{(l-1,0)}  s \partial_c \mathcal{Y}_{l-1}^0$, so that finally $w_{10}(\theta , \varphi)=0$.

\section{Application of the generalized Lamb solution to Marangoni flows}
\label{secappli}

Bringing everything together, we finally reach the conclusion that the generalized exterior solution for the Stokes flow is given by Equations~(\ref{componentgen}) and~(\ref{velocitygen}), supplemented with Equations~(\ref{vl1gen}) and~(\ref{vthetal10gen}) for $l=1$. We now apply these results to  thermocapillary flows. 

\subsection{Back to the first encounter}

In Section~\ref{secthermocap}, we discussed the thermocapillary flow due to a point source at the liquid-air interface. At vanishing P\'eclet number, the temperature field is found to decay as $r^{-1}$, see Equation~(\ref{tempfield}). The Marangoni boundary condition then suggests that the velocity field assumes the same functional form:
$v_r(r,\theta) = \Phi_r(\theta)/r$ and $v_{\theta}(r,\theta) = \Phi_{\theta}(\theta)/r$.
Keeping only the terms proportional to $r^{-1}$ in the generalized exterior solution, we can directly write
\begin{subequations}
\begin{align}
&  \Phi_r(\theta)  = \pi_{(10)} c + \rho_{(-1,0)}   \ , \\
&  \Phi_{\theta}(\theta)  = -  \frac{\pi_{(10)}}{2} s -  \rho_{(-1,0)} \sqrt{ \frac{1-c}{1+c} }   \ .
\end{align}
\end{subequations}
If we moreover enforce the condition of vanishing normal velocity at the interface, \textit{i.e.} $v_{\theta}(r,\frac{\pi}{2})=0$, then the integration constants are related through 
$\pi_{(10)}=-2\rho_{(-1,0)}$. The solution then reads
\begin{subequations}
\begin{align}
&  \Phi_r(\theta)  = \rho_{(-1,0)} (1-2c)   \ , \\
&  \Phi_{\theta}(\theta)  = \rho_{(-1,0)} \frac{cs}{1+c}    \ ,
\end{align}
\end{subequations}
which is precisely the result given in Equation~(\ref{bratukhin}). The remaining constant can be set by enforcing the stress continuity condition~(\ref{bcmar}). We get
$\rho_{(-1,0)}=Q\gamma_T/(4\pi \kappa \eta)$, as expected.

\subsection{Thermocapillary flow with dipolar symmetry}

As a second example, we consider the self-propulsion of a heated particle at the water-air interface. This question has recently  attracted much attention, both  theoretically and experimentally~\cite{girotLangmuir2016,wurgerJFM2014,masoudJFM2014}. The particle acts as a point-like heat source. It is assumed that the temperature profile in the liquid phase has a dipolar symmetry
\begin{equation}
T(\mathbf{r}) = T_0 + \frac{Q}{2\pi \kappa} \left( \frac{1}{r} + \frac{\mathbf{b} \cdot \mathbf{r} }{r^3} \right)  \ ,  \label{dipotemp}
\end{equation}
with $\mathbf{b}=b\mathbf{e}_x$.
The resulting thermocapillary flow is solution of the Stokes Equations~(\ref{stokes})--(\ref{incomp})  together with the Marangoni boundary condition~(\ref{bcmar}).
The explicit expression for the velocity field was obtained in Ref.~\cite{wurgerJFM2014} after  tedious calculations.  
Here, we can get it directly from the generalized Lamb solution. According to the superposition principle, the velocity field can be written as 
\begin{equation}
\mathbf{v} ( \mathbf{r}) = \mathbf{v}_{10} ( \mathbf{r}) + b \mathbf{v}_{21} ( \mathbf{r}) \ ,
\end{equation}
where the axisymmetric term $\mathbf{v}_{10}$ has been discussed in the previous paragraph. It is then straightforward to get the dipolar contribution
\begin{subequations}
\begin{align}
u_{2,1} (c) & = \pi_{(2,1)} \mathcal{P}_2^1(c) + \rho_{(0,1)}\mathcal{P}_0^1(c)    
 = \frac{1}{2s} \left[ 2\rho_{(0,1)} + \left( \pi_{(2,1)} -2 \rho_{(0,1)}\right) c - \pi_{(2,1)}  c^3 \right]  \ , \\
v_{2,1} (c) & = s \rho_{(0,1)} \partial_c \mathcal{P}_0^1 + \frac{i}{s} \sigma_{(1,1)} \mathcal{P}_1^1(c)  
 = \frac{i \sigma_{(1,1)}}{2}  - \rho_{(0,1)} \frac{1}{1+c} \ ,  \\
 w_{2,1} (c) & = -\frac{i}{s} \rho_{(0,1)}\mathcal{P}_0^1(c)  + s \sigma_{(1,1)} \partial_c \mathcal{P}_1^1   = 
-i \left[  \rho_{(0,1)} \frac{1}{1+c} -  \frac{i \sigma_{(1,1)}}{2} c \right] \ .
\end{align}
\end{subequations}
This solution matches exactly with the one that was calculated    in Ref.~\cite{wurgerJFM2014} [see Equation~(4.6)].

\section{Conclusions}
\label{conclusion}

To summarize, we have derived a generalization of Lamb solution for the Stokes flow in a semi-infinite domain.
Equations~(\ref{componentgen})--(\ref{vthetal10gen}) are the main results of this work.
This new solution is relevant for instance when the liquid phase is bounded by a flat interface, or in the case of a hemispherical liquid drop moving on an inhomogeneous surface. The range of applications is therefore relatively wide. 

We emphasize that, although our results look quite similar to the original Lamb solution, it presents several features that are rather unconventional. First of all, the degree $l$ is not limited to integers but is in general a real number. Second, the order $m$ is not limited to the interval $-l \leq m \leq l$ but can take values larger than the degree $l$. These peculiarities are directly related to the fact that the flow is restricted to a semi-infinite region.

The expression of the velocity components~(\ref{componentgen})--(\ref{vthetal10gen}) can be used directly for any problem with hemispherical geometry, exactly as one would proceed with Lamb solution. As shown in Section~\ref{secappli}, it is especially well suited for interfacial phenomena involving the Marangoni effect. We thus expect that this formalism might shed a new light on several important issues in interfacial science, including the spreading of surfactant from a point source or the actuation of Marangoni surfers.

\acknowledgments{The authors wish to thank  A. W\"urger for many useful and fruitful discussions.}

\appendix

\section{From Legendre to Gauss hypergeometric equation}
\label{appgauss}

In this appendix, we follow Reference~\cite{seabornbook} to construct the general solutions of the associated Legendre equation in a semi-infinite region
\begin{equation}
(1-c^2) \frac{\d^2 g}{\d c^2}-2c\frac{\d g}{\d c}+\left[ l(l+1) -\frac{m^2}{1-c^2} \right] g(c) = 0 \ .
\label{legendreeqapp}
\end{equation}
First, it  should be noticed that  Equation~(\ref{legendreeqapp}) admits three regular singular points at $c=1$, $c=-1$ and $c = + \infty$, respectively. As a consequence, it is always possible to recast Equation~(\ref{legendreeqapp}) into Gauss hypergeometric equation~\cite{seabornbook}
\begin{equation}
x(1-x) y''(x) + \left[ \gamma - (\alpha+\beta+1) x \right] y'(x) - \alpha \beta y(x) = 0 \ ,
\label{hypergeoeqapp}
\end{equation}
where $\alpha$, $\beta$ and $\gamma$ are three real (yet unknown) parameters. A fundamental solution of Equation~(\ref{hypergeoeqapp}) is  given by the hypergeometric function $F(\alpha,\beta;\gamma;x)$, which is defined as
\begin{equation} 
F(\alpha,\beta;\gamma;x) = \sum_{n=0}^{\infty} \frac{(\alpha)_n(\beta)_n}{(\gamma)_n} \frac{x^n}{n!} = \frac{\Gamma(\gamma)}{\Gamma (\alpha) \Gamma (\beta) }
\sum_{n=0}^{\infty} \frac{\Gamma (\alpha + n) \Gamma (\beta + n) }{\Gamma(\gamma + n)} \frac{x^n}{n!}  \ ,
\label{defhyperapp}
\end{equation}
with $(q)_n= q(q+1)\ldots (q+n-1)$  the Pochhammer symbol [$(q)_0=1$], and  $\Gamma(x)=\int_0^{\infty} t^{x-1} e^{-t} d t$  the gamma function. The hypergeometric function verifies the symmetry relation
\begin{equation} 
F(\alpha,\beta;\gamma;x) = F(\beta, \alpha;\gamma;x)  \ .
\end{equation}
The radius of convergence of the series is $R=1$. The series reduces to a polynomial of degree $l \in \mathbb{N}$ when either $\alpha$ or $\beta$ is equal to $-l$.

The solution of Equation~(\ref{hypergeoeqapp}) being known, the issue is  to find the mathematical transformation that relates Equations~(\ref{legendreeqapp}) and~(\ref{hypergeoeqapp}).
To this aim, we make the substitution $g(c) = \chi(x) y(x)$, with $\chi(x)$ an auxiliary function such that $y(x)$ is solution of Equation~(\ref{hypergeoeqapp}). We also define the new variable $x=(1-c)/2$. Inserting  in Equation~(\ref{legendreeqapp}), one gets an intermediate  differential equation for $y$
\begin{align}
x(1-x) & \chi  y''  + \Big[ 2x(1-x) \chi' + (1-2x) \chi\Big] y'   \nonumber \\
& + \left\{  x(1-x) \chi'' + (1-2x) \chi' + \left( l(l+1) - \frac{m^2}{4x(1-x)} \right) \chi \right\} y = 0 \ .
\label{eqchiapp}
\end{align}
On comparing the second- and first-order terms with those in Equation~(\ref{hypergeoeqapp}), one can infer that the term in square brackets must be proportional to $\chi$. Without loss of generality, one can always absorb any overall constant factor in the definition of $y$. We thus obtain a first equation for $\chi$
\begin{equation}
2x(1-x) \chi' + (1-2x) \chi = \left[ \gamma - (\alpha+\beta+1) x \right] \chi \ ,
\label{eqchi1app}
\end{equation}
whose solution reads
\begin{equation}
\chi(x) = x^{(\gamma-1)/2}(1-x)^{(\alpha+\beta-\gamma)/2}\ .
\label{solchiapp}
\end{equation}
Then, comparison of the zeroth-order terms in Equations~(\ref{hypergeoeqapp}) and~(\ref{eqchiapp}) imposes that the term in curly brackets must be proportional to $\chi$ as well
\begin{equation}
 x(1-x) \chi'' + (1-2x) \chi' + \left\{ l(l+1) - \frac{m^2}{4x(1-x)} \right\} \chi = - \alpha \beta \chi \ .
\label{eqchi2app}
\end{equation}
Inserting the solution~(\ref{solchiapp}) in Equation~(\ref{eqchi2app}), we can deduce that
\begin{equation}
\gamma -1 = \epsilon_1 \vert m \vert \ , \quad \alpha + \beta - \gamma = \epsilon_2 \vert m \vert \ , \quad \text{and} \quad \alpha \beta = -l(l+1) \ .
\label{parametersgaussapp}
\end{equation}
Here we define $\epsilon_i = \pm 1$. The problem therefore admits four independent solutions. For the discussion, it is convenient to switch back to the original variable~$c$.
\begin{itemize}
\item If $\epsilon_1 = \epsilon_2= +1$, then the auxiliary function reads $\chi(c) = \frac{1}{4} (1-c^2)^{\vert m \vert /2}$. This is the canonical definition usually adopted in the literature (see, \textit{e.g.}, Reference~\cite{seabornbook}) in order to account for the regularity condition over the whole interval $c \in [-1,1]$. 
\item If $\epsilon_1 =+1$ and $\epsilon_2= -1$, then we get
\begin{equation}
\chi(c) = \left( \frac{1-c}{1+c} \right)^{\vert m \vert /2} \ .
\label{reschiapp}
\end{equation}
This function has a singularity at the south pole $c=-1$, but is regular otherwise. It is thus relevant in the upper half-space.
\item If $\epsilon_1 =-1$ and $\epsilon_2=+1$, we obtain $\chi(c) = \left[ (1+c)/(1-c) \right]^{\vert m \vert /2}$.
The singularity is now located at the north pole $c=1$. This configuration being complementary to the previous one, it is suitable in the lower half-space.
\item If $\epsilon_1 = \epsilon_2= -1$, the auxiliary function $\chi(c) = 4 (1-c^2)^{-\vert m \vert /2}$
is singular at both poles. It is therefore inappropriate with regard to the semi-infinite domains discussed in this work.
\end{itemize} 
In the upper half-space, the relevant choice is then $\epsilon_1 =+1$ and $\epsilon_2= -1$. It  follows from Equation~(\ref{parametersgaussapp}) that the three Gauss parameters  are 
\begin{equation}
\alpha = - l \ , \quad \beta = l+1 \ , \quad \text{and} \quad \gamma = \vert m \vert +1 \ .
\end{equation}
Finally, the  solutions of Equation~(\ref{legendreeqapp})  are the generalized Legendre functions, defined for  $c\in[0,1]$ by
\begin{equation} 
\mathcal{P}_l^m(c)= \left( \frac{1-c}{1+c} \right)^{\vert m \vert /2} F\left(-l,l+1;\vert m \vert+1;\frac{1-c}{2}\right)   \ .
\label{defplmapp}
\end{equation}
In the complementary interval  $c\in[-1,0]$, the solutions are  obtained thanks to the substitution $c \leftrightarrow -c$.
It can be noted that the generalized Legendre functions satisfy the condition
\begin{equation}
\mathcal{P}_{-(l+1)}^m(c) = \mathcal{P}_l^m(c)  \ .
\end{equation}
They also verify the following properties: (i) they are eigenfunctions of the differential operator $\mathcal{L}$  
\begin{equation}
\mathcal{L} \mathcal{P}_l^m(c)  \doteq \left[ (1-c^2) \partial^2_c -2c \partial_c - \frac{m^2}{(1-c^2)} \right] \mathcal{P}_l^m(c) = -l(l+1) \mathcal{P}_l^m(c)  \ .
\end{equation}
Note that one also has
\begin{equation}
\mathcal{L}\left[c  \mathcal{P}_l^m \right] = -(l^2+l+2) c \mathcal{P}_l^m + 2 (1-c^2) \frac{\d \mathcal{P}_l^m}{\d c}    \ .
\end{equation}
(ii) They follow the recurrence relations
\begin{subequations}
\begin{align}
& \left(1-c^2 \right) \frac{\d \mathcal{P}_l^m}{\d c} = \left( l- \vert m \vert \right)\mathcal{P}_{l-1}^m(c) -lc \mathcal{P}_l^m(c) \ ,  \label{rec1}\\
& \left(1-c^2 \right) \frac{\d \mathcal{P}_l^m}{\d c} = - \left( l+ \vert m \vert +1 \right)\mathcal{P}_{l+1}^m(c) +(l+1)c \mathcal{P}_l^m(c) \ . \label{rec2}
\end{align}
\end{subequations}

\section{Solutions of the Stokes problem in a semi-infinite region}
\label{appl0}

The aim of this second appendix is to derive explicitly the solution of the Stokes Equations~(\ref{stokes})--({\ref{incomp}) in the semi-infinite region defined by $c = \cos \theta \geq 0$. 
The parameter~$s= \sin \theta=\sqrt{1-c^2}$ is always positive since $\theta \in [0, \pi]$.
Note also that $\partial_{\theta} = -s \partial_c$ and $\d s / \d c = - c/s$.
In this representation, Equation~(\ref{incomp}) reads
\begin{equation}
\label{contapp}
\frac{1}{r^2} \partial_r \left( r^2 v_r \right) - \frac{1}{r} \partial_c (s v_{\theta} ) + \frac{1}{sr} \partial_{\varphi}  v_{\varphi} = 0 \ .
\end{equation}
The projection of the Stokes equations on the spherical basis $\left( \mathbf{e}_r, \mathbf{e}_{\theta}, \mathbf{e}_{\varphi} \right)$ leads to a set of partial differential equations that are listed for instance in~\cite{landaubook}. We then look for solutions using the Anzats suggested in Equations~(\ref{componentgen}).

\subsection{Pressure field}

The pressure $p(\mathbf{r})$ satisfies Laplace equation $\Delta p=0$, where we define the scalar Laplacian
\begin{equation}
\Delta \psi = r^{-1} \partial^2_r \left( r \psi \right) + r^{-2} \partial_c \left(s^2 \partial_c \psi \right) + (rs)^{-2} \partial^2_{\varphi}  \psi \ .
\end{equation}
Inserting the Ansatz $p(r,\theta,\varphi) = r^{-(l+1)} p_{lm}(c)e^{im\varphi}$, one obtains the Legendre differential equation  for each mode $p_{lm}(c)$
\begin{equation}
(1-c^2) \frac{\d^2 p_{lm}}{\d c^2}-2c\frac{\d p_{lm}}{\d c}+\left[ l(l+1) -\frac{m^2}{1-c^2} \right] p_{lm}(c) = 0 \ .
\end{equation}
Following the discussion of Appendix~\ref{appgauss}, it is straightforward to conclude that 
\begin{equation}
\label{plm}
p_{lm}(c) =  \frac{2(2l-1)}{l+1} \pi_{(l,m)} \mathcal{P}_l^m(c) \ ,
\end{equation}
with $\pi_{(lm)}$ the integration constant. Here, the prefactor $2(2l-1)/(l+1)$ is introduced for later convenience.

\subsection{Radial component of the velocity} 

The radial projection of the Stokes Equation~(\ref{stokes})  reads 
\begin{equation}
\Delta v_r - \frac{2}{r^2} v_r + \frac{2}{r^2} \partial_c \left( s v_{\theta} \right) - \frac{2}{s r^2} \partial_{\varphi} v_{\varphi} = \frac{1}{\eta} \partial_r p \ .
\end{equation}
Then, eliminating $v_{\theta}$ and $v_{\varphi}$ thanks to Equation~(\ref{contapp}), one gets 
\begin{equation}
\Delta v_r - \frac{2}{r^2} v_r + \frac{2}{r^3} \partial_r \left( r^2 v_r \right) = \frac{1}{\eta} \partial_r p \ .
\end{equation}
We then assume the functional form $v_r(r,\theta,\varphi) = r^{-l} u_{lm}(c)e^{im\varphi}$, so that each mode satisfies the following equation
\begin{equation}
(1-c^2) \frac{\d^2 u_{lm}}{\d c^2}-2c\frac{\d u_{lm}}{\d c}+\left[(l-2)(l-1) -\frac{m^2}{1-c^2} \right] u_{lm}(c) = -2(2l-1) p_{lm}(c)  \ .
\end{equation}
One recognizes  on the left-hand side the Legendre equation of degree $l-2$: the homogeneous solution is thus $u_{lm}^{(h)}(c) \propto \mathcal{P}_{l-2}^m(c)$. 
Moreover, since $p_{lm}(c) \propto \mathcal{P}_l^m(c)$, a particular solution 
can be searched under the form $u_{lm}^{(p)}(c) = \alpha \mathcal{P}_{lm}(c)$, with $\alpha$ a constant. It is then straightforward to get $\alpha=\pi_{(lm)}$, so that finally
\begin{equation}
\label{ulm}
u_{lm}(c) = \pi_{(l,m)} \mathcal{P}_l^m(c) + \rho_{(l-2,m)} \mathcal{P}_{l-2}^m(c) \ .
\end{equation}

\subsection{Polar component of the velocity}

The projection of the Stokes Equation~(\ref{stokes}) on the polar direction leads to
\begin{equation}
\Delta v_{\theta} - \frac{1}{s^2r^2} v_{\theta} - \frac{2s}{r^2} \partial_c v_r - \frac{2c}{s^2 r^2} \partial_{\varphi} v_{\varphi} = - \frac{s}{\eta r} \partial_c p \ .
\end{equation}
Defining $\tilde{v}_{\theta} = s v_{\theta}$ and eliminating $v_{\varphi}$ thanks to Equation~(\ref{contapp}), it can be rewritten as
\begin{equation}
s^2 \partial^2_{c} \tilde{v}_{\theta} -2c \partial_c \tilde{v}_{\theta} + \frac{1}{s^2} \partial^2_{\varphi} \tilde{v}_{\theta}+ r \partial^2_r \left( r \tilde{v}_{\theta} \right)
=   - \frac{s^2 r}{\eta} \partial_c p  + 2s^2 \partial_c v_r - \frac{2c}{r} \partial_r \left( r^2 v_r \right) \ .
\label{stokespol}
\end{equation}
We then proceed along the same lines  and look for solutions of the form  $v_{\theta} = r^{-l} v_{lm}(c)e^{im\varphi}$. Setting $\tilde{v}_{lm} = s v_{lm}$ yields to
\begin{align}
\label{eqvtheta}
 (1-c^2) \frac{\d^2 \tilde{v}_{lm}}{\d c^2}-2c\frac{\d \tilde{v}_{lm}}{\d c}+  & \left[l(l-1) -  \frac{m^2}{1-c^2} \right]  \tilde{v}_{lm}(c) \nonumber  \\
&  = -(1-c^2) \frac{ \d p_{lm}}{\d c}+2(1-c^2) \frac{ \d u_{lm}}{\d c} +2(l-2)c u_{lm}(c) \ .
\end{align}
The left-hand side (LHS) corresponds to Legendre equation of degree $l-1$, so that the homogeneous solution is  $\tilde{v}^{(h)}_{lm}(c) \propto \mathcal{P}_{l-1}^m(c)$. Coming back to the original function 
we get $v^{(h)}_{lm}(c) \propto \mathcal{P}_{l-1}^m(c)/s$, which happens to be singular at $c=1$ when $m=0$. But since one can always search for a particular solution that is regular at this point (except for the case $l=1$, see below), the homogeneous solution has to be regular as well. To remove the singularity, one is naturally led to set
\begin{equation}
v^{(h)}_{lm}(c) = im \sigma_{(l-1,m)} \mathcal{P}_{l-1}^m(c)/s  \ .
\end{equation}
Note that the homogeneous term can also be written as $v^{(h)}_{lm}(c)  e^{im\varphi} = \sigma_{(l-1,m)} \partial_{\varphi} \mathcal{Y}_{l-1}^m/s$, with 
$\mathcal{Y}_l^m(c,\varphi) = \mathcal{P}_l^m(c)e^{im\varphi} $, which is the  form  generally used in Lamb solution.

We next focus on the inhomogeneous equation. According to the previous results, the right-hand side (RHS) is a linear combination of $\mathcal{P}_l^m$ and $\mathcal{P}_{l-2}^m$. Due to the linearity of Equation~(\ref{eqvtheta}), the particular solution can be searched under the form $\tilde{v}_{lm}^{(p)}=\tilde{v}_{lm}^{(p1)}+\tilde{v}_{lm}^{(p2)}$, with  
\begin{equation}
\tilde{v}_{lm}^{(p1)}  =  K_1 (1-c^2) \frac{\d \mathcal{P}_l^m}{\d c}  \ , \quad \text{and} \quad \tilde{v}_{lm}^{(p2)} =  K_2 (1-c^2) \frac{\d \mathcal{P}_{l-2}^m}{\d c} \ .
\end{equation}
The two constants $K_1\propto \pi_{(lm)}$ and $K_2 \propto \rho_{(lm)}$ have yet to be determined. Thanks to the recurrence relations 
of the Legendre functions Equation~(\ref{rec1}), we can write
\begin{equation}
\tilde{v}_{lm}^{(p1)}  = K_1\left[ (l-\vert m\vert ) \mathcal{P}_{l-1}^m -lc \mathcal{P}_{l}^m \right] \ ,
\end{equation}
and 
\begin{equation}
\tilde{v}_{lm}^{(p2)} = K_2 \left[ (l-1)c \mathcal{P}_{l-2}^m -(l+m-1) \mathcal{P}_{l-1}^m \right] \ .
\end{equation}
Inserting into the LHS of Equation~(\ref{eqvtheta}), one finds LHS=LHS$_1$+LHS$_2$, with
\begin{subequations}
\begin{align}
& \text{LHS}_1 =  -2l K_1  \left[(1-c^2) \frac{\d \mathcal{P}_l^m}{\d c} -(l+1)c \mathcal{P}_l^m \right]  \label{lhs1} \ , \\
& \text{LHS}_2 = 2(l-1) K_2  \left[(1-c^2) \frac{\d \mathcal{P}_{l-2}^m}{\d c} +(l-2)c \mathcal{P}_{l-2}^m \right] \label{lhs2} \ .
\end{align}
\end{subequations}
From the solution (\ref{plm}) for $p_{lm}$ and (\ref{ulm}) for $u_{lm}$, the RHS of Equation~(\ref{eqvtheta}) can also be written as RHS=RHS$_1$+RHS$_2$, with
\begin{subequations}
\begin{align}
& \text{RHS}_1 =  -2\frac{l-2}{l+1} \pi_{(lm)}  \left[(1-c^2) \frac{\d \mathcal{P}_l^m}{\d c} -(l+1)c \mathcal{P}_l^m \right]  \label{rhs1} \ , \\
& \text{RHS}_2 = 2 \rho_{(l-2,m)}  \left[(1-c^2) \frac{\d \mathcal{P}_{l-2}^m}{\d c} +(l-2)c \mathcal{P}_{l-2}^m \right]  \ . \label{rhs2}
\end{align}
\end{subequations}
Comparing (\ref{lhs1}) and (\ref{rhs1}) on the one hand, and (\ref{lhs2}) and (\ref{rhs2}) on the other hand, one finally obtains
\begin{equation}
K_1 =  \frac{l-2}{l(l+1)} \pi_{(lm)}   \ , \quad \text{and} \quad K_2 = \frac{1}{l-1} \rho_{(l-2,m)}  \ .
\end{equation}
We can then draw the  conclusion that
\begin{equation}
\label{vlm}
v_{lm}(c) =  s\left[ \frac{l-2}{l(l+1)} \pi_{(lm)} \frac{\d \mathcal{P}_l^m}{\d c} + \frac{\rho_{(l-2,m)}}{l-1}   \frac{\d \mathcal{P}_{l-2}^m}{\d c} \right] + \frac{im}{s} \sigma_{(l-1,m)} \mathcal{P}_{l-1}^m(c)  \ .
\end{equation}

\subsection{Azimuthal component of the velocity}

Although one can proceed in the same way for the azimuthal projection, it is more convenient to deduce the component $v_{\varphi}$ directly  from Equation~(\ref{contapp}).
Defining $v_{\varphi} = r^{-l} w_{lm}(c)e^{im\varphi}$, the modes $w_{lm}$ are given by
\begin{equation}
w_{lm}(c) =  -\frac{is}{m} \left[ \frac{\d \left( sv_{lm}\right)}{\d c} + (l-2) u_{lm}(c) \right]  \ .
\label{contwlm}
\end{equation}
Using~(\ref{ulm}), (\ref{vlm}) and the definition of Legendre equation, we get after some algebra
\begin{equation}
\label{wlm}
w_{lm}(c) =  -\frac{im}{s} \left[ \frac{(l-2)}{l(l+1)} \pi_{(lm)} \mathcal{P}_l^m(c) + \frac{\rho_{(l-2,m)}}{l-1}  \mathcal{P}_{l-2}^m(c) \right] + s \sigma_{(l-1,m)} 
\frac{\d \mathcal{P}_{l-1}^m}{\d c}  \ . 
\end{equation}

\subsection{Singular cases}

It can be noticed that the expressions of $v_{lm}$ and $w_{lm}$ are not defined for $l=1$, see Equations~(\ref{vlm}) and~(\ref{wlm}).
These singular cases can be handled as follows. Using  the recurrence relation~(\ref{rec2}), Equation~(\ref{rhs2}) can be rewritten as
\begin{equation}
\text{RHS}_2 = 2 \rho_{(l-2,m)} (l-\vert m \vert -2) \mathcal{P}_{l-3}^m   \ ,
\end{equation}
so that the corresponding particular solution can be searched under the form $\tilde{v}^{(p2)}_{lm}=  K'_2 \mathcal{P}_{l-3}^m$. After some algebra one gets
\begin{equation}
K'_2 =   \left( \frac{l-\vert m \vert -2}{2l-3} \right)  \rho_{(l-2,m)}  \ .
\end{equation}
This expression is now regular for $l=1$. We can thus write
\begin{equation}
\label{v1m}
v_{1m}(c) = -  \frac{\pi_{(1m)}}{2}   s  \partial_c  \mathcal{P}_1^m +  
(1+\vert m \vert ) \rho_{(-1,m)} \frac{\mathcal{P}_{1}^m}{s} +   \sigma_{(0,m)}   \frac{im}{s} \mathcal{P}_{0}^m  \ , \quad m \neq 0 \ .
\end{equation}
But the latter relation has still to be regularized for $m=0$. To do so, we set $\tilde{\sigma}_{(00)} = \left[im \sigma_{(0,m)}\right]_{m=0}$ and  replace in Equation~(\ref{v1m}) the generalized Legendre functions by their explicit expression
\begin{equation}
v_{10}(c)  =  \frac{ \pi_{(10)} }{2s} \left[ c^2 +  2 \frac{ \rho_{(-1,0)} }{  \pi_{(10)}  } c  + \left( 2 \frac{ \tilde{\sigma}_{(0,0)} }{ \pi_{(10)}} -1 \right)  \right] 
\end{equation}
For this expression to be regular, $c=1$ must be a root of the term in squared brackets, \textit{i.e.}
\begin{equation}
\big[ \ldots \big]= \left( c-1 \right) \left[ c+  \left( \frac{2\rho_{(-1,0)}}{\pi_{(10)}}  +1 \right) \right] \ .
\end{equation}
This finally sets the value $\tilde{\sigma}_{(0,0)}= - \rho_{(-1,0)}$, so that
\begin{equation}
v_{10}(c) =   - \frac{1}{2}\pi_{(10)} \sqrt{1-c^2} -   \rho_{(-1,0)} \sqrt{\frac{1-c}{1+c}}   \ . 
\end{equation}


\end{document}